\documentclass[10pt,conference]{IEEEtran}

\usepackage{graphicx,epsf,psfrag}
\usepackage{amsmath,amssymb}
\usepackage{amsfonts}
\usepackage{mathrsfs}
\usepackage{etoolbox}
\usepackage{cite}

\usepackage{latexsym}

\newcommand{\beq}{\begin{equation}}
\newcommand{\eeq}{\end{equation}}
\newcommand{\be}{\begin{equation}}
\newcommand{\ee}{\end{equation}}
\newcommand{\eps}{\epsilon}

\newcommand{\bi}{\begin{itemize}}
\newcommand{\ei}{\end{itemize}}

\newcommand{\calA}{\mathcal{A}}

\newcommand{\calD}{\mathcal{D}}
\newcommand{\calE}{\mathcal{E}}

\newcommand{\calP}{\mathcal{P}}

\newcommand{\calS}{\mathcal{S}}

\newcommand{\calX}{\mathcal{X}}
\newcommand{\calY}{\mathcal{Y}}

\newcommand{\bbE}{\mathbb{E}}

\newcommand{\bbP}{\mathbb{P}}

\newcommand{\bbR}{\mathbb{R}}

\DeclareMathAlphabet{\mathbsf}{OT1}{cmss}{bx}{n}
\DeclareMathAlphabet{\mathssf}{OT1}{cmss}{m}{sl}

\newcommand{\rvQ}{\mathsf{Q}}

\DeclareSymbolFont{bsfletters}{OT1}{cmss}{bx}{n}  
\DeclareSymbolFont{ssfletters}{OT1}{cmss}{m}{n}
\DeclareMathSymbol{\bsfGamma}{0}{bsfletters}{'000}
\DeclareMathSymbol{\ssfGamma}{0}{ssfletters}{'000}
\DeclareMathSymbol{\bsfDelta}{0}{bsfletters}{'001}
\DeclareMathSymbol{\ssfDelta}{0}{ssfletters}{'001}
\DeclareMathSymbol{\bsfTheta}{0}{bsfletters}{'002}
\DeclareMathSymbol{\ssfTheta}{0}{ssfletters}{'002}
\DeclareMathSymbol{\bsfLambda}{0}{bsfletters}{'003}
\DeclareMathSymbol{\ssfLambda}{0}{ssfletters}{'003}
\DeclareMathSymbol{\bsfXi}{0}{bsfletters}{'004}
\DeclareMathSymbol{\ssfXi}{0}{ssfletters}{'004}
\DeclareMathSymbol{\bsfPi}{0}{bsfletters}{'005}
\DeclareMathSymbol{\ssfPi}{0}{ssfletters}{'005}
\DeclareMathSymbol{\bsfSigma}{0}{bsfletters}{'006}
\DeclareMathSymbol{\ssfSigma}{0}{ssfletters}{'006}
\DeclareMathSymbol{\bsfUpsilon}{0}{bsfletters}{'007}
\DeclareMathSymbol{\ssfUpsilon}{0}{ssfletters}{'007}
\DeclareMathSymbol{\bsfPhi}{0}{bsfletters}{'010}
\DeclareMathSymbol{\ssfPhi}{0}{ssfletters}{'010}
\DeclareMathSymbol{\bsfPsi}{0}{bsfletters}{'011}
\DeclareMathSymbol{\ssfPsi}{0}{ssfletters}{'011}
\DeclareMathSymbol{\bsfOmega}{0}{bsfletters}{'012}
\DeclareMathSymbol{\ssfOmega}{0}{ssfletters}{'012}

\newcommand{\tilQ}{\tilde{Q}}

\newcommand{\bars}{\bar{s}}

\newcommand{\barx}{\bar{x}}

\newcommand{\barX}{\bar{X}}

\DeclareMathOperator*{\esssup}{ess\,sup}

\DeclareMathOperator{\var}{Var}

\ifcsmacro{theorem}{}{
\newtheorem{theorem}{Theorem}
\newtheorem{lemma}[theorem]{Lemma}

\newtheorem{corollary}[theorem]{Corollary}
 
\newtheorem{example}{Example} 
\newtheorem{remark}{Remark}

}

\newcommand{\qednew}{\nobreak \ifvmode \relax \else
      \ifdim\lastskip<1.5em \hskip-\lastskip
      \hskip1.5em plus0em minus0.5em \fi \nobreak
      \vrule height0.75em width0.5em depth0.25em\fi}

\allowdisplaybreaks

\title{Finite Blocklength and Dispersion Bounds for the Arbitrarily-Varying Channel}

\author{Oliver Kosut and J{\"o}rg Kliewer
\thanks{O.~Kosut is with the School of Electrical, Computer and Energy Engineering, Arizona State University, Tempe, AZ 85287 USA (email: \hbox{okosut@asu.edu}).}
\thanks{J. Kliewer is with the Department of Electrical and Computer Engineering, New Jersey Institute of Technology, Newark, NJ 07102 USA (email: \hbox{jkliewer@njit.edu}).}
\thanks{This material is based upon work supported by the National Science
  Foundation under Grant No. CCF-1439465, CCF-1440014, CNS-1526547, CCF-1453718.}
}

\IEEEoverridecommandlockouts

\begin{document}

\maketitle

\begin{abstract}
Finite blocklength and second-order (dispersion) results are presented for the arbitrarily-varying channel (AVC), a classical model wherein an adversary can transmit arbitrary signals into the channel. A novel finite blocklength achievability bound is presented, roughly analogous to the random coding union bound for non-adversarial channels. This finite blocklength bound, along with a known converse bound, are used to derive bounds on the dispersion of discrete memoryless AVCs without shared randomness, and with cost constraints on the input and the state. These bounds are tight for many channels of interest, including the binary symmetric AVC. However, the bounds are not tight if the deterministic and random code capacities differ.
\end{abstract}

\newcommand{\T}{T_{\eps}^{(n)}}
\renewcommand{\t}{T^{(n)}}

\section{Introduction}

Active, malicious adversaries represent a potential threat against modern communication systems. This is particularly true of wireless systems, in which the inherently open nature of the communication medium allows for an intelligent jammer to transmit a damaging signal. The arbitrarily-varying channel (AVC) is a classical information-theoretic model that captures an active adversary in a point-to-point setting. Classical work on the AVC characterized the capacity with and without shared randomness between the encoder and decoder, and in which the input and state (or adversarial signal) are subject to cost constraints.

In this paper, we present finite blocklength and second-order results for the AVC under average probability of error and \emph{without} shared randomness, including cases with cost constraints. We introduce a novel finite blocklength achievability bound, which is a strengthened form of the achievability bound used in \cite{Csiszar1988} to derive the AVC capacity without shared randomness. We further show that in some cases, this achievability bound is strong enough to achieve both the capacity and the dispersion of discrete memoryless AVCs. The \emph{dispersion} characterizes the asymptotic second-order behavior of a channel subject to a fixed probability of error constraint. Analysis of this sort dates back to Strassen \cite{Strassen}, and has seen significant interest in recent years, particularly since \cite{Polyanskiy2010}. The dispersion of the compound channel, which is closely related to the AVC---in fact, they are indistinguishable in the single-shot setting (see Remark~\ref{remark:cc})---was derived for discrete memoryless channels in \cite{Polyanskiy2013}. We found the dispersion of AVCs with shared randomness between encoder and decoder in our prior work \cite{Kosut2017a}, although this result did not extend to channels with cost constraints. In the present paper, we provide the exact dispersion of discrete memoryless AVCs without shared randomness, and with or without cost constraints, provided certain conditions are satisfied. These conditions are satisfied for some channels of interest, such as binary symmetric AVCs, but not others, including parts of the parameter space for the binary adding AVC.

\section{Preliminaries}

\subsection{Notation}

Given a set $\calX$, let $\calP(\calX)$ be the set of random distributions with alphabet $\calX$. For some $P\in\calP(\calX)$, we write $X\sim P$ to mean that $X$ is a random variable drawn from distribution $P$. The probability measure is denoted $\bbP$, and the expectation operator is denoted $\bbE$; the underlying distribution will be specified in context. Given a function $g:\calX\to\bbR$ and a real number $\Gamma$, let $\calP(\calX,\Gamma)$ be the set of distributions $P\in\calP(\calX)$ where $\bbE g(X)\le \Gamma$ if $X\sim P$. The underlying function $g$ will be understood from the context. Also let
\be
\calX^n(\Gamma)=\textstyle\{x^n\in\calX^n:\sum_{i=1}^n g(x_i)\le n\Gamma\}.
\ee
For alphabet $\calS$, function $\ell:\calS\to \bbR$ and real number $\Lambda$, we define $\calP(\calS,\Lambda)$ and $\calS^n(\Lambda)$ similarly. Let $\calP(\calY|\calX)$ be the set of conditional distributions $P_{Y|X}$ where $P_{Y|X}(\cdot|x)\in\calP(\calY)$ for all $x\in\calX$. For any $P_X\in\calP(\calX)$ and $P_{Y|X}\in\calP(\calY|\calX)$, we write $P_XP_{Y|X}\in\calP(\calY)$ where
\be
(P_XP_{Y|X})(y)=\sum_{x\in\calX} P_X(x)P_{Y|X}(y|x).
\ee
Similarly, given $P_S\in\calP(\calS)$ and $W\in\calP(\calY|\calX\times\calS)$, let $P_SW\in\calP(\calY|\calX)$ be given by
\be
(P_SW)(y|x)=\sum_{s\in\calS} P_S(s)W(y|x,s).
\ee
Note that $P_XP_SW\in\calP(\calY)$ is now also well defined. Given $P_X\in\calP(\calX)$ or $P_{Y|X}(\calY|\calX)$, and any positive integer $n$ we write their stationary-memoryless extensions as $P_X^n\in\calP(\calX^n)$ and $P_{Y|X}^n\in\calP(\calY^n|\calX^n)$ where
\[
P_X^n(x^n)=\prod_{i=1}^n P_X(x_i),~~ P_{Y|X}^n(y^n|x^n)=\prod_{i=1}^n P_{Y|X}(y_i|x_i).
\]
Given a sequence $x^n\in\calX^n$, its type is given by
\be
Q_{x^n}(x)=\frac{1}{n}|\{i:x_i=x\}|.
\ee
Let $\calP_n(\calX)$ be the set of all types of sequences in $\calX^n$. For $P\in\calP_n(\calX)$, let $T(P)$ be the type class of $P$; i.e., the set of sequences $x^n\in\calX^n$ with $Q_{x^n}=P$. Also, for $P\in\calP_n(\calX)$, let $U_{P_X}$ be the uniform distribution over type class $T(P)$. For any integer $M$, we write $[M]=\{1,\ldots,M\}$. Finally, $\log$ and $\exp$ are assumed to have base $2$.

\subsection{Problem Description}

We first describe a single-shot AVC model, with the input, state, and output alphabets having arbitrary structure, and the channel itself represented by an arbitrary conditional probability measure. Subsequently, we specialize the model to the $n$-length stationary memoryless case. 

A single-shot AVC is given by the tuple $(\calX,\calS,W(y|x,s),\calY)$ where $W\in\calP(\calY|\calX\times\calS)$. An $(M,\eps)$ code is given by an encoding function
$
\phi:[M]\to \calX
$
and a decoding function
$
\psi:\calY\to[M]
$
where for any $s\in\calS$, the average probability of error is at most $\eps$; i.e. 
\be
\sup_{s\in\calS}\, \frac{1}{M}\sum_{m=1}^M W(\psi^{-1}(m)^c|\phi(m),s)\le\eps
\ee
where $\psi^{-1}(m)^c$ is the set of $y\in\calY$ such that $\psi(y)\ne m$. Let $M^\star(\eps)$ be the largest integer $M$ for which there exists an $(M,\eps)$ code.

Given cost functions $g:\calX\to\bbR$ and $\ell:\calS\to\bbR$, an $n$-length cost-constrained AVC is given by the tuple 
\be
(\calX^n(\Lambda),\calS^n(\Gamma),W^n(y^n|x^n,s^n),\calY^n).
\ee
where $\Lambda,\Gamma$ are real numbers. An $(M,n,\eps)$ code consists of a code for this channel with $M$ messages and probability of error $\eps$. Define $M^\star(n,\eps)$ similarly.

\begin{remark}\label{remark:cc}
While in this paper we are primarily interested in the AVC, the above single-shot model is indistinguishable from a compound channel model, which differs from an AVC only in that the state must be held constant across the coding block, a distinction that only makes sense in the $n$-length setting. In fact, our finite blocklength achievability bound Thm.~\ref{thm:avc_rcu}, which applies in the general single-shot setting, may be considered as an achievable bound for the compound channel as well as the AVC.
\end{remark}

\section{Finite Blocklength Achievability Bound}

The following theorem is our new achievability bound for the AVC. As we will illustrate below, this bound is analogous to the random coding union (RCU) bound for non-state channels, as derived in \cite{Polyanskiy2010}.

\begin{theorem}\label{thm:avc_rcu}
Fix $P_X$, and let $Z(x,\barx,y)\in\{0,1\}$ be a test such that
\be\label{eq:unique_restriction}
Z(x,\barx,y)Z(\barx,x,y)=0\text{ for all }x,\barx\in\calX,\,y\in\calY
\ee
and let $\calA\subseteq\calX\times\calY$. For each $s\in\calS$, let $(X,\barX,Y_s)\sim P_X(x) P_X(\barx) W(y|x,s)$. There exists an $(M,\eps)$ code such that
\begin{align}
\eps&\le
\max_s\,
\bbP((X,Y_s)\notin\calA)\nonumber
\\&+(2\log e) M\,\bbP(Z(X,\barX,Y_s)=0,\,(X,Y_s)\in\calA)\nonumber
\\&+\esssup\, 2\log(3|\calS|)\, \bbP(Z(X,\barX,Y_s)=0,\,(X,Y_s)\in\calA|\barX)\nonumber
\\&+\sqrt{\frac{2\ln(3|\calS|)}{M}}.\label{eq:avc_rcu}
\end{align}
\end{theorem}

The test $Z$ can be viewed as a test for whether $x$ is more likely than $\barx$ to be the transmitted codeword, given that $y$ has been received by the decoder. Specifically, the proof of Thm.~\ref{thm:avc_rcu} uses the following decoding rule for codebook $\{c_1,\ldots,c_M\}$:
\emph{Given output $y$, decode to message $i$ if $Z(c_i,c_j,y)=1$ for all $j\ne i$. If there is no such message, declare an error.} Note that condition \eqref{eq:unique_restriction} ensures that two messages cannot simultaneously satisfy this criterion. The set $\calA$ can be thought of as a jointly typical set of input-output pairs.

\begin{remark}
From Thm.~\ref{thm:avc_rcu}, one can recover a bound similar to the RCU bound of \cite{Polyanskiy2010} as follows. Given a channel without state (i.e., $|\calS|=1$), we may choose
\be
Z(x,\barx,y)=\mathbf{1}\big(\imath(x;y)>\imath(\barx;y)\big)
\ee
where $\imath(x;y)$ is the information density. This test clearly satisfies \eqref{eq:unique_restriction}. One can now see that the optimal choice for $\calA$ to minimize the first two terms in \eqref{eq:avc_rcu} is
\be
\calA=\{(x,y):(2\log e)M\,\bbP(\imath(\barX,y)\ge \imath(x;y))\le 1\}.
\ee
Thus the first two terms in \eqref{eq:avc_rcu} become
\be
\bbE \min\big\{1,(2\log e)M\,\bbP\big(\imath(\barX,Y)\ge \imath(X;Y)\big|X,Y\big)\big\}.
\ee
This expression is nearly identical to the standard RCU bound, except that $M-1$ has been replaced by $(2\log e)M$. This difference constitutes less than 2 bits. Furthermore, the last two terms in \eqref{eq:avc_rcu} are vanishingly small.
\end{remark}

The proof of Thm.~\ref{thm:avc_rcu} relies on the following lemma, which is a sharpened version of \cite[Lemma A1]{Csiszar1988}. The lemma is a Chernoff bound that holds even for variables that are not i.i.d., provided they have a bounded conditional expectation.

\begin{lemma}\label{lemma:large_deviation}
Let $X_1,\ldots,X_M$ be random variables and let $f_i(x_1,\ldots,x_i)$ be a set of $M$ functions where
\be
\bbE[f_i(X_1,\ldots,X_i)|X_1,\ldots,X_{i-1}]\le \mu\quad\text{a.s.}
\ee
and
$
f_i(X_1,\ldots,X_i)\in[0,\gamma]
$ a.s.
Then for all $t\in[\mu,\gamma]$,
\begin{multline}
\bbP\left(\frac{1}{M} \sum_{i=1}^M f_i(X_1,\ldots,X_i)>t\right)
\\< \min\left\{2^{-M(\frac{t-\mu \log_2 e}{\gamma})},
e^{-2M(\frac{t-\mu}{\gamma})^2}\right\}.\label{eq:chernoff_bound_loose}
\end{multline}
\end{lemma}
\begin{IEEEproof}
We first prove that 
\be
\bbP\left(\frac{1}{M} \sum_{i=1}^M f_i(X_1,\ldots,X_i)>t\right)
<\exp\left\{-M D\left(\frac{t}{\gamma}\bigg\|\frac{\mu}{\gamma}\right)\right\}\label{eq:chernoff_bound}
\ee
where $D(p\|q)$ is the relative entropy between Bernoulli random variables. If $\frac{1}{M}\sum_{i=1}^M f_i(X_1,\ldots,X_i)=\mu$ a.s., then \eqref{eq:chernoff_bound} holds trivially for all $t\in[\mu,\gamma]$. Otherwise, we assume $\gamma=1$; the result immediately generalizes to other values. For any $\lambda>0$ we have
\begin{align}
&\bbP\left(\frac{1}{M} \sum_{i=1}^M f_i(X_1,\ldots,X_i)>t\right)\nonumber
\\&= \bbP\left(\exp\left\{\lambda\sum_i f_i(X_1,\ldots,X_i)\right\}>\exp\{\lambda M t\}\right)
\\&< \exp\{-\lambda M t\} \bbE \exp\left\{\lambda \sum_i f_i(X_1,\ldots,X_i)\right\}\label{eq:markov_application}
\\&\le \exp\{-\lambda M t\} \bbE \prod_i \big[1+(\exp\lambda-1)f_i(X_1,\ldots,,X_i)\big]
\\&\le  \exp\{-\lambda M t\}\big[1+(\exp\lambda-1)\mu\big]^M
\\&=  \exp\big\{-M\big[\lambda t -\log\left(1+(\exp\lambda-1)\mu\right)\big]\big\}
\end{align}
where \eqref{eq:markov_application} follows from Markov's inequality, and the strict inequality holds because $\sum_i f_i(X_1,\ldots,X_i)$ is not constant, and non-negative. To prove \eqref{eq:chernoff_bound}, we note that
\be\label{eq:lambda_D}
\sup_{\lambda>0} \lambda t -\log\left(1+(\exp\lambda-1)\mu\right)
=D(t\|\mu).
\ee
To prove \eqref{eq:chernoff_bound_loose}, we lower bound the relative entropy in two ways. First, by choosing $\lambda=\log 2$ in \eqref{eq:lambda_D}, we have
\be
D(t\|\mu)\ge t\log 2-\log(1+\mu)\ge t\log 2-\mu\log e.
\ee
This proves the first bound in \eqref{eq:chernoff_bound_loose}. To prove the second, note that 
\be
\frac{d}{dt} D(t\|\mu)\big|_{t=\mu}=0
\ee
and
\be
\frac{d^2}{dt^2} D(t\|\mu)=\log e\left(\frac{1}{t}+\frac{1}{1-t}\right)\ge 4\log e.
\ee
Therefore
\be
D(t\|\mu)\ge 2(t-\mu)^2\log e.
\ee
\end{IEEEproof}

\begin{IEEEproof}[Proof of Thm.~\ref{thm:avc_rcu}]
Applying the decoding rule described above, given codebook $\{c_1,\ldots,c_M\}$ and state $s$, the average probability of error is
\begin{multline}\label{eq:init_pe_bd}
P_e(c_1,\ldots,c_M|s)\\=\frac{1}{M}\sum_i \bbP\big(Z(c_i,c_j,Y_s)=0\text{ for some }j\ne i\big|X=c_i\big).
\end{multline}
Recall that $\calA$ is some subset of $\calX\times\calY$ representing a jointly typical set. We may upper bound the probability of error by
\begin{align}
&P_e(c_1,\ldots,c_M|s)
 \le 
\frac{1}{M} \sum_i \bigg[ 
\bbP\Big((c_i,Y_s)\notin\calA\nonumber
\\&\text{ or }Z(c_i,c_j,Y_s)=0\text{ for some }j<i\Big)\nonumber
\\&+\bbP\Big((c_i,Y_s)\in\calA,\,
Z(c_i,c_j,Y_s)=0\text{ for some }j>i\Big)\bigg].\label{eq:pe_bd1}
\end{align}
Let $C_1,\ldots,C_M$ be independent random variables, each drawn from $P_X$. We proceed to show that with some positive probability, $P_e(C_1,\ldots,C_M|s)$ exceeds the quantity in the RHS of \eqref{eq:avc_rcu} for all $s\in\calS$. Let
\begin{align}
q(\barx,s)&=\bbP(Z(X,\barx,Y_s)=0|(X,Y_s)\in\calA).
\end{align}
Now let $f_i(c_1,\ldots,c_i|s)=0$ if $\sum_{j<i} q(c_j,s)>Mt_{1s}$ (where $t_{1s}$ is a constant to be determined), and otherwise
\begin{multline}\label{eq:fi_def}
f_i(c_1,\ldots,c_i|s)=\bbP\Big((c_i,Y_s)\notin\calA\\ \text{ or }Z(c_i,c_j,Y_s)=0\text{ for some }j<i\Big).
\end{multline}
Similarly, let $g_i(c_i,\ldots,c_M|s)=0$ if $\sum_{j>i} q(c_j,s)>Mt_{1s}$, and otherwise
\begin{multline}\label{eq:gi_def}
g_i(c_i,\ldots,c_M|s)=\bbP\Big((c_i,Y_s)\in\calA,\,Z(c_i,c_j,Y_s)=0\\ \text{ for some }j>i\Big).
\end{multline}
We now define three classes of error events (again $t_{2s},t_{3s}$ are to be determined):
\begin{align}
\calE_{1s}&=\left\{\frac{1}{M}\sum_i q(C_i,s)>t_{1s}\right\},\\
\calE_{2s}&=\left\{\frac{1}{M}\sum_i f_i(C_1,\ldots,C_i|s)>t_{2s}\right\},\\
\calE_{3s}&=\left\{\frac{1}{M}\sum_i g_i(C_i,\ldots,C_M|s)>t_{3s}\right\}.
\end{align}
Note that if $\calE_{1s}$ does not occur, then RHS of \eqref{eq:pe_bd1} is equal to $\frac{1}{M}\sum_i[f_i(c_1,\ldots,c_M|s)+g_i(c_1,\ldots,c_M|s)]$. We proceed to find constants $t_{1s},t_{2s},t_{3s}$ such that the probability that each of these events is less than $(3|\calS|)^{-1}$, thus proving that there exists at least one code that does not fall into any of these events. Define
\begin{align}
\alpha_s&=\bbE q(\barX,s)=\bbP(Z(X,\barX,Y_s)=0|(X,Y_s)\in\calA),\\
\gamma_s&=\esssup q(\barX,s).\label{eq:gamma_s_def}
\end{align}
Note that in \eqref{eq:gamma_s_def}, the essential supremum corresponds to a supremum over the support set of $\barX$. If we choose
\be
t_{1s}=\alpha_s \log e+\frac{\gamma_s \log(3|\calS|)}{M}.
\ee
then by Lemma~\ref{lemma:large_deviation}
\begin{align}
\bbP(\calE_{1s})&=
\bbP\left(\frac{1}{M} \sum_i q(C_i,s)>t_{1s}\right)
< 2^{-M(\frac{t_{1s}-\alpha_s \log e}{\gamma_s})}\nonumber
\\&=(3|\calS|)^{-1}.
\end{align}
If $\sum_{j<i} q(c_j,s)\le Mt_{1s}$ then for any fixed $c_1,\ldots,c_{i-1}$,
\begin{align}
&\bbE f_i(c_1,\ldots,c_{i-1},C_i)
\\&\le\bbP((X,Y_s)\notin\calA)+\sum_{j<i} \bbP(Z(X,c_j,Y_s)=0,\,(X,Y_s)\in\calA)
\\&=\bbP((X,Y_s)\notin\calA)+\bbP((X,Y_s)\in\calA)\sum_{j<i} q(c_j,s)
\\&\le \bbP((X,Y_s)\notin\calA)+\bbP((X,Y_s)\in\calA)M t_{1s}.\label{eq:f_bd1}
\end{align}
Moreover, the upper bound in \eqref{eq:f_bd1} holds for all $(c_1,\ldots,c_{i-1})$, since when $\sum_{j<i} q(c_j,s)> Mt_{1s}$ the function is identically zero. If we choose
\be
t_{2s}=\bbP((X,Y_s)\notin\calA)+\bbP((X,Y_s)\in\calA)M t_{1s}+\sqrt{\frac{\ln(3|\calS|)}{2M}}.
\ee
then by Lemma~\ref{lemma:large_deviation} and the fact that $f_i\in[0,1]$,
\begin{align}
\bbP(\calE_{2s})&=\bbP\left(\frac{1}{M}\sum_i f_i(C_1,\ldots,C_i)>t_{2s}\right)
\\&< e^{-2M(t_{2s}-\bbP((X,Y_s)\notin\calA)+\bbP((X,Y_s)\in\calA)M t_{1s})^2}
\\&=(3|\calS|)^{-1}.
\end{align}
By a similar argument, $\bbP(\calE_{3s})<(3|\calS|)^{-1}$ if
\be
t_{3s}=\bbP((X,Y_s)\in\calA)M t_{1s}+\sqrt{\frac{\ln(3|\calS|)}{2M}}.
\ee
Therefore, there exists a codebook $\{c_1,\ldots,c_M\}$ falling into no error events for any $s$. In particular, since $\calE_{1s}$ does not occur, the functions $f_i,g_i$ are equal to the expressions in \eqref{eq:fi_def}--\eqref{eq:gi_def} (rather than zero), so we may rewrite the RHS of \eqref{eq:init_pe_bd} to conclude that for all $s$
\begin{align}
&P_e(c_1,\ldots,c_M|s)\nonumber
\\&\le \frac{1}{M}\sum_i \big[f_i(c_1,\ldots,c_i|s)+g_i(c_i,\ldots,c_M|s)\big]
\\&\le t_{2s}+t_{3s}
\\&=
\bbP((X,Y_s)\notin\calA)
+2\bbP((X,Y_s)\in\calA)M t_{1s}+\sqrt{\frac{2\ln(3|\calS|)}{M}}
\\&=
\bbP((X,Y_s)\notin\calA)+\bbP((X,Y_s)\in\calA)\big[2(\log e)M \alpha_s\nonumber
\\&\qquad+2\gamma_s \log(3|\calS|)\big]
+\sqrt{\frac{2\ln(3|\calS|)}{M}}
\\&=\bbP((X,Y_s)\notin\calA)\nonumber
\\&\quad+(2\log e) M\,\bbP(Z(X,\barX,Y_s)=0,\,(X,Y_s)\in\calA)\nonumber
\\&\quad+\esssup\, 2\log(3|\calS|)\, \bbP(Z(X,\barX,Y_s)=0,\,(X,Y_s)\in\calA|\barX)\nonumber
\\&\quad+\sqrt{\frac{2\ln(3|\calS|)}{M}}.
\end{align}
\end{IEEEproof}

\section{Dispersion Bounds}

Consider an $n$-length cost-constrained AVC with finite alphabets, given by the single-letter conditional distribution $W(y|x,s)$. Given $P_X\in\calP(\calX)$ and $P_S\in\calP(\calS)$, let $(X,S,Y)\sim P_X(x)P_S(s)W(y|x,s)$. Now we define the following information quantities:
\begin{align}
\imath(x;y)&=\log\frac{(P_SW)(y|x)}{(P_XP_SW)(y)},\\
I(P_X,P_{Y|X})&=\bbE \imath_{P_{Y|X}\|(P_X P_{Y|X})}(X;Y),\\
\tilde\imath(x;s;y)&=\imath(x;y)-\bbE(\imath(X;Y)|X=x)\nonumber
\\&-\bbE(\imath(X;Y)|S=s)
+I(P_X,P_SW),\\
V(P_X,P_S,W)&=\bbE\, \tilde\imath(X;S;Y)^2,
\\T(P_X,P_S,W)&=\bbE\, |\tilde\imath(X;S;Y)|^3.
\end{align}

For any $P_X\in\calP(\calX)$, let
\be
\Lambda_0(P_X)=\min_{P_{S|X}}\ \sum_{x\in\calX,\,s\in\calS} P_X(x)P_{S|X}(s|x)\ell(s)
\ee
where the minimum is over distributions $P_{S|X}\in\calP(\calS|\calX)$ such that, for all $x,x'$ where $P_X(x),P_X(x')>0$,
\be\label{eq:symmetrizing_distributions}
\sum_{s} P_{S|X}(s|x) W(y|x',s)=\sum_s P_{S|X}(s|x')W(y|x,s),
\ee
and $\Lambda_0(P_X)=\infty$ if there is no distribution satisfying \eqref{eq:symmetrizing_distributions}. An AVC is said to be \emph{symmetrizable} if $\Lambda_0(P_X)\le\Lambda$ for all $P_X\in\calP(\calX,\Gamma)$, in which case the capacity is zero. For non-symmetrizable AVCs, the capacity was found in \cite{Csiszar1988} to be
\be\label{eq:AVC_capacity}
C=\max_{\substack{P_X\in\calP(\calX,\Gamma):\\ \Lambda_0(P_X)\ge\Lambda}}\ \min_{P_S\in\calP(\calS,\Lambda)}\,I(P_X,P_SW).
\ee
Note that the feasible sets for both the maximum and minimum in \eqref{eq:AVC_capacity} are convex sets. Moreover, mutual information is concave in the input distribution and convex in the channel distribution, so the maximum and minimum in \eqref{eq:AVC_capacity} can be exchanged without changing the value. We may define $\Pi_X(\Gamma)$ and $\Pi_S(\Lambda)$ to be the sets of optimal distributions for $P_X$ and $P_S$ respectively. Let
\be
V_+=\min_{P_X\in\Pi_X(\Gamma)}\ \max_{P_S\in\Pi_S(\Lambda)}\, V(P_X,P_S,W)\label{eq:Vplus}
\ee
For a cost-constrained AVC, the \emph{random code capacity}---defined as the capacity when the encoder and decoder have access to an unlimited amount of shared randomness, unknown to the adversary---is given by \cite{Csiszar1988a}
\be\label{eq:random_code_capacity}
C_r=\max_{P_X\in\calP(\calX,\Gamma)}\ \min_{P_S\in\calP(\calS,\Lambda)}\,I(P_X,P_SW).
\ee
Let $\Pi_X^{(r)}(\Gamma)$ and $\Pi_S^{(r)}(\Lambda)$ be the set of optimal distributions for $P_X$ and $P_S$ in \eqref{eq:random_code_capacity}. Let
\be
V_-=\max_{P_S\in\Pi_S^{(r)}(\Lambda)}\  \min_{P_X\in\Pi_X^{(r)}(\Gamma)}\,V(P_X,P_S,W).\label{eq:Vminus}
\ee
Let $\rvQ$ be the complementary CDF of the standard Gaussian distribution, and $\rvQ^{-1}$ its inverse.

The following theorems give upper and lower bounds on the normal approximation for discrete-memoryless AVCs. 

\begin{theorem}\label{thm:dmc_converse}
Consider an $n$-length, cost-constrained AVC. For any $\eps\in(0,1/2)$, 
\begin{multline}
\log M^\star(n,\eps)\le nC_r-\sqrt{nV_-}\,\rvQ^{-1}(\eps)\\+(|\calX|+|\calS|-{\textstyle\frac{3}{2}})\log n+O(1).
\end{multline}
\end{theorem}

\begin{theorem}\label{thm:dmc_achievability}
Consider a cost-constrained AVC for which there exists a distribution $P_X^\star\in\Pi_X(\Gamma)$ that achieves the minimum in \eqref{eq:Vplus}  such that $\Lambda_0(P_X^\star)>\Lambda$. Then for any $\eps\in(0,1/2)$, 
\begin{multline}
\log M^\star(n,\eps)\ge nC-\sqrt{nV_+}\,\rvQ^{-1}(\eps)\\-(|\calX|+|\calS|-{\textstyle\frac{3}{2}})\log n-O(1).
\end{multline}
\end{theorem}

While our bounds do not match even to first order when the random code capacity exceeds the capacity, the following corollary gives a sufficient condition for the bounds to hold up to second order.

\begin{corollary}\label{corollary:sufficient_condition}
Consider a cost-constrained non-symmetrizable AVC such that: (i) there exists a distribution $P_X^\star\in\Pi^{(r)}(\Lambda)$ where $\Lambda_0(P_X^\star)>\Lambda$, and (ii) at least one of the sets $\Pi_X(\Gamma)$ and $\Pi_S(\Lambda)$ contain only a single element. Then $C_r=C$, $V_+=V_-$, and for any $\eps\in(0,1/2)$,
\be
\log M^\star(n,\eps)=nC-\sqrt{nV_+}\,\rvQ^{-1}(\eps)+O(\log n).
\ee
\end{corollary}

We now consider two examples, illustrating cases in which the sufficient condition in Corollary~\ref{corollary:sufficient_condition} does or does not hold. The capacity of both of these examples was originally found in \cite{Csiszar1988}.

\begin{example}[Binary symmetric AVC]
Let $\calX,\calY,\calS=\{0,1\}$, and $W(y|x,s)=1$ if $y=x\oplus s$, where $\oplus$ is addition modulo 2. Let $g(x)=x$ and $\ell(s)=s$. If $P_X=[1-p,p]$, then $\Lambda_0(P_X)=\min\{p,1-p\}$. Thus, the channel is symmetrizable if $\Lambda\ge\min\{\Gamma,1/2\}$. Otherwise, the capacity and the random code capacity are both $H(\Gamma(1-\Lambda)+(1-\Gamma)\Lambda)-H(\Lambda)$, where $H(\cdot)$ is the binary entropy function. Moreover, the optimal input and state distributions in both \eqref{eq:AVC_capacity} and \eqref{eq:random_code_capacity} are unique, so this channel satisfies the conditions of Corollary~\ref{corollary:sufficient_condition}. The dispersion is given by
\[
V_+=\begin{cases} 4\Gamma(1-\Gamma)\Lambda(1-\Lambda)\log^2\frac{\Lambda+\Gamma-2\Lambda\Gamma}{1-\Lambda-\Gamma+2\Lambda\Gamma},&\Gamma\le 1/2\\ 0&\Gamma>1/2.\end{cases}
\]
Of particular note is that, even though the capacity is the same as a non-adversarial binary symmetric channel with crossover probability $\Lambda$, the dispersion is strictly smaller.
\end{example}

\begin{example}[Binary adding AVC]
Let $\calX,\calS=\{0,1\}$, $\calY=\{0,1,2\}$, and $W(y|x,s)=1$ if $y=x+s$, where we are using real-valued addition. Again let $g(x)=x$ and $\ell(s)=s$. If $P_X=[1-p,p]$, then $\Lambda_0(P_X)=p$. Thus, the channel is symmetrizable if $\Gamma\le \Lambda$. If $\Gamma>\Lambda$ and $\Lambda\le 1/2$, then the capacity and the random code capacity are equal (although with no simple closed form), and moreover the optimal input and state distributions are unique, so the sufficient conditions of Corollary~\ref{corollary:sufficient_condition} are satisfied. However, if $\Gamma>\Lambda>1/2$, then the capacity and random code capacity differ, in which case our results do not give tight bounds on the dispersion.
\end{example}

Before we prove Thms.~\ref{thm:dmc_converse} and \ref{thm:dmc_achievability}, we state several lemmas. The first provides a necessary continuity result.

\begin{lemma}\label{lemma:eta_set}
Assume $\calX$, $\calS$, and $\calY$ are finite sets, and that $P_X(x)>0$ for all $x\in\calX$. Let $\calD_\eta$ be the set of joint distributions $Q_{XX'SY}$ such that $Q_S\in\calP(\calS,\Lambda)$ and
\be
D(Q_{XX'SY}\|P_X\times Q_{X'S}\times W)\le\eta
\ee
where 
\be(P_X\times Q_{X'S}\times W)(x,x',s,y)=P_X(x) Q_{X'S}(x',s) W(y|x,s).\ee 
Let
\begin{multline}\label{eq:eta_star_def}
\eta^\star=\inf\{\eta: Q_{XX'SY}\in\calD_\eta\text{ and }Q_{X'XS'Y}\in\calD_\eta
\\\text{ for some }Q_{XX'SSY}\}.
\end{multline}
If $\Lambda_0(P_X)>\Lambda$, then $\eta^\star>0$.
\end{lemma}
\begin{IEEEproof}
We prove the contrapositive: namely, if $\eta^\star=0$, then $\Lambda_0(P_X)\le\Lambda$. Assuming $\eta^\star=0$, then for all $\eta>0$, there exists a distribution $Q_{XX'SS'Y}$ such that $Q_{XX'SY}\in\calD_\eta$ and $Q_{X'XS'Y}\in\calD_\eta$, and where $Q_S,Q_{S'}\in\calP(\calS,\Lambda)$. Thus, by continuity of relative entropy on discrete alphabets and compactness of the set of distributions $Q_{XX'SS'Y}$, there exists a distribution $Q_{XX'SS'Y}$ such that $Q_{XX'SS'}\in\calD_0$ and $Q_{X'XS'Y}\in\calD_0$, where again $Q_S,Q_{S'}\in\calP(\calS,\Lambda)$. That is,
\begin{align}
Q_{XX'SY}(x,x',s,y)&=P_X(x) Q_{X'S}(x',s) W(y|x,s),\\
Q_{X'XS'Y}(x',x,s',y)&= P_{X'}(x') Q_{XS'}(x,s') W(y|x',s').
\end{align}
Note that $Q_X=Q_{X'}=P_X$, and
\begin{multline}
\sum_{s}P_X(x) P_X(x')Q_{S|X'}(s|x') W(y|x,s) 
\\= \sum_{s'}P_X(x') P_X(x) Q_{S'|X}(s'|x) W(y|x',s').
\end{multline}
Thus, for all $x,x'$ in the support of $P_X$,
\be
\sum_{s}Q_{S|X'}(s|x') W(y|x,s) = \sum_{s'} Q_{S'|X}(s'|x) W(y|x',s').
\ee
If we define $\tilQ_{S|X}=\frac{1}{2}(Q_{S|X'}+Q_{S'|X})$, then we may switch places and average to find that for all $x,x'$ in the support of $P_X$,
\be
\sum_{s} \tilQ_{S|X}(s|x')W(y|x,s)=\sum_s \tilQ_{S|X}(s|x)W(y|x',s)
\ee
which is precisely the condition for a symmetrizing distribution in \eqref{eq:symmetrizing_distributions}. Therefore, since $Q_S,Q_{S'}\in\calP(\calS,\Lambda)$,
\begin{align}
\Lambda_0(P_X)
&\le \sum_{x,s} P_X(x) \tilQ_{S|X}(s|x)\ell(s)
\\&=\sum_s \frac{1}{2}(Q_S(s)+Q_{S'}(s))
\\&\le \Lambda.
\end{align}
\end{IEEEproof}

The following lemma is a slight restatement of \cite[Thm. 3]{Scarlett2015a}, a Berry-Esseen-type result for interacting constant-composition distributions, which was itself derived from a result on Latin hypercube sampling in \cite{Loh1996}. This lemma is key to deriving the dispersion of the AVC under input and state constraints, just as it was in \cite{Scarlett2015a} to derive the dispersion of constant-composition codebooks for the multiple-access channel.

\begin{lemma}\label{lemma:latin_square}
Given $P_X\in\calP_n(\calX)$ and $P_S\in\calP_n(\calS)$, let
\be
(X^n,S^n,Y^n)\sim U_{P_X}(x^n)\,U_{P_S}(s^n)\,W^n(y^n|x^n,s^n).
\ee
Let
$
Z_n=\sum_{i=1}^n \imath(X_i;Y_i)
$
and let $\Sigma_n=\frac{1}{n}\var(Z_n)$. If $V(P_X,P_S,W)>0$, then for all $\gamma$,
\be
\left|\bbP\left(\frac{Z_n-\bbE Z_n}{\sqrt{n\Sigma_n}}>\gamma\right)-\rvQ(\gamma)\right|\le \frac{K\,T(P_X,P_S,W)}{\Sigma_n^{3/2}\sqrt{n}}
\ee
where $K$ is an absolute constant. Moreover, 
\be\label{eq:Sigma_n_bound}
0\le\Sigma_n-V(P_X,P_S,W)\le\frac{3}{n-1}\var(\imath(X;Y)).
\ee
\end{lemma}

\begin{IEEEproof}[Proof of Thm.~\ref{thm:dmc_converse}]
Let $P_S^\star\in\Pi_S^{(r)}(\Lambda)$ achieve the maximum in \eqref{eq:Vminus}. Let $P_S\in\calP_n(\calS)\cap\calP(\calS,\Lambda)$ be such that $\|P_S-P_S^\star\|_\infty\le 1/n$. The adversary may randomly choose the state sequence from $U_{P_S}$, inducing the non-adversarial channel $U_{P_S}W^n$. Thus, an upper bound on the achievable rate for this non-adversarial channel is also an upper bound on the underlying AVC. From here on, we only consider this non-adversarial channel. We first bound the number of messages in constant composition codes. Specifically, for any $P_X\in\calP_n(\calX,\Gamma)$, consider an $(M,n,\eps)$ code with codewords entirely in $T_{P_X}$. Applying the finite blocklength non-adversarial converse bound \cite[Proposition 4.4]{Tan2014a}, for any $\delta>0$,
\begin{align}
&\eps+\delta\nonumber
\\&\ge \sup_{Q_{Y^n}}\,\max_{x^n\in T_{P_X}}\,\bbP\left\{\log \frac{(U_{P_S}W^n)(Y^n|x^n)}{Q_{Y^n}(Y^n)}\le \log (M\delta)\right\}\label{eq:upper1}
\\&\ge \max_{x^n\in T_{P_X}}\,\bbP\left\{\log \frac{(U_{P_S}W^n)(Y^n|x^n)}{(P_XP_SW)^n(Y^n)}\le \log (M\delta)\right\}\label{eq:upper2}
\\&= \bbP\left\{\log \frac{(U_{P_S}W^n)(Y^n|X^n)}{(P_XP_SW)^n(Y^n)}\le \log (M\delta)\right\}\label{eq:upper3}
\\&\ge \bbP\left\{\log \frac{(P_SW)^n(Y^n|X^n)}{(P_XP_SW)^n(Y^n)}\le \log (M\delta)-\log|\calP_n(\calS)|\right\}\label{eq:upper4}
\end{align}
where in \eqref{eq:upper1}--\eqref{eq:upper2}, $Y^n\sim (U_{P_S}W^n)(y^n|x^n)$, whereas in \eqref{eq:upper3}--\eqref{eq:upper4}, $(X^n,Y^n)\sim U_{P_X}(x^n)(U_{P_S}W^n)(y^n|x^n)$; in \eqref{eq:upper2} we have chosen $Q_{Y^n}=(P_XP_SW)^n$; \eqref{eq:upper3} holds since the quantity in \eqref{eq:upper2} depends only on the type of $x^n$; and \eqref{eq:upper4} holds because $U_{P_S}(s^n)\le P_S^n(s^n)|\calP_n(\calS)|$ for all $s^n$. From \eqref{eq:upper4}, we may apply an argument identical to that of \cite[Thm. 49]{Polyanskiy2010}, with Lemma~\ref{lemma:latin_square} in place of the Berry-Esseen theorem, to conclude that for any $(M,n,\eps)$ code, $\log M$ is at most
\begin{multline}\label{eq:PS_bound}
\max_{P_X\in\calP(\calX,\Lambda)} nI(P_X;P_SW)
-\sqrt{nV(P_X,P_S,W)} \rvQ^{-1}(\eps)
\\ +(|\calX|+|\calS|-{\textstyle\frac{3}{2}})\log n+O(1).
\end{multline}
Let
\begin{align}
C(P_S)&=\max_{P_X\in\calP(\calX,\Lambda)}\ I(P_X;P_SW)\label{eq:CPS}\\
V_{\min}(P_S)&=\min_{P_X}\ V(P_X,P_S,W)\label{eq:Vmin}
\end{align}
where the minimum in \eqref{eq:Vmin} is over distributions that achieve the maximum in \eqref{eq:CPS}. Applying \cite[Lemmas 63 and 64]{Polyanskiy2010}, we may further upper bound \eqref{eq:PS_bound} by
\begin{align}
\log M&\le nC(P_S)-\sqrt{nV_{\min}(P_S)} \,\rvQ^{-1}(\eps)\nonumber
\\&\qquad+(|\calX|+|\calS|-{\textstyle\frac{3}{2}})\log n+O(1)
\\&\le nC(P_S^\star)-\sqrt{nV_{\min}(P_S^\star)} \,\rvQ^{-1}(\eps)\nonumber
\\&\qquad+(|\calX|+|\calS|-{\textstyle\frac{3}{2}})\log n+O(1)\label{eq:PS_PSstar}
\end{align}
where \eqref{eq:PS_PSstar} holds since $\|P_S-P_S^\star\|_\infty\le 1/n$, so replacing $P_S$ by $P_S^\star$ changes the value by no more than $O(1)$. Noting that $C(P_S^\star)=C_r$ and $V_{\min}(P_S^\star)=V_-$ completes the proof.
\end{IEEEproof}

\begin{IEEEproof}[Proof of Thm.~\ref{thm:dmc_achievability}]
Let $P_X^\star\in\Pi_X(\Gamma)$ achieve the minimum in \eqref{eq:Vplus}, with $\Lambda_0(P_X^\star)>\Lambda$, the existence of which is assumed in the statement of the theorem. Let $P_X\in\calP_n(\calX)\cap\calP(\calX,\Gamma)$ be such that $\|P_S-P_S^\star\|_\infty\le 1/n$. By continuity, for sufficiently large $n$ we have $\Lambda_0(P_X)>\Lambda$. Let
\begin{multline}
\calA=\bigg\{(x^n,y^n):\log\frac{(P_SW)^n(y^n|x^n)}{(U_{P_X}P_S^nW^n)(y^n)}\ge\gamma\\ \text{for some }P_S\in\calP_n(\calS)\bigg\}
\end{multline}
where we define with hindsight
\be\label{eq:gamma_def}
\gamma=\log\big[\sqrt{n}\,|\calP_n(\calS)\,M\big].
\ee 
By Lemma~\ref{lemma:eta_set} we have $\eta^\star>0$, so we may fix $0<\eta<\eta^\star$. Define a test given by $Z(x^n,\barx^n,y^n)=1$ if $(x^n,y^n)\in\calA$, and either $(\barx^n,y^n)\notin\calA$ or there exists $s^n$ such that
\be\label{eq:Z_def_condition}
Q_{x^n,\barx^n,s^n,y^n}\in \calD_\eta.
\ee
Note that if $Z(x^n,\barx^n,y^n)Z(\barx^n,x^n,y^n)=1$, then $(x^n,y^n)\in\calA$, $(\barx^n,y^n)\in\calA$, and there exist $s^n,\bars^n$ such that
\be\label{eq:two_types}
Q_{x^n,\barx^n,s^n,y^n}\in\calD_\eta,\qquad Q_{\barx^n,x^n,\bars^n,y^n}\in\calD_\eta.
\ee
However, since $\eta<\eta^\star$, by the definition of $\eta^\star$ in \eqref{eq:eta_star_def}, the two conditions in \eqref{eq:two_types} cannot occur simultaneously. Therefore
\be\label{eq:Zn_zero}
Z(x^n,\barx^n,y^n)Z(\barx^n,x^n,y^n)=0 \text{ for all }x^n,\barx^n,y^n.
\ee

Having proved \eqref{eq:Zn_zero}, we may apply Thm.~\ref{thm:avc_rcu} with $X^n\sim U_{P_X}$ to find that there exists an $(M,n,\eps)$ code where\footnote{Recall that $Y^n_{s^n}$ indicates the channel output sequence with state sequence $s^n$.}
\begin{align}
\eps&\le \max_{s^n\in\calS^n(\Lambda)}\, \bbP((X^n,Y^n_{s^n})\notin\calA)\nonumber
\\&+(2\log e)M\,\bbP(Z(X^n,\barX^n,Y^n_{s^n})=0,\,(X^n,Y^n_{s^n})\in\calA)\nonumber
\\&+\max_{\barx^n}\, 2\log(3n|\calS|)\,\bbP(Z(X^n,\barx^n,Y^n_{s^n})=0,(X^n,Y^n_{s^n})\in\calA)\nonumber
\\&+\sqrt{\frac{2\ln(3n|\calS|)}{M}}.\label{eq:dmc_terms}
\end{align}
We may bound the first term in \eqref{eq:dmc_terms} by
\begin{align}
&\bbP((X^n,Y^n_{s^n})\notin\calA)\nonumber
\\&=\bbP\left(\log\frac{(P_SW)^n(Y^n_{s^n}|X^n)}{(U_{P_X}P_S^nW^n)(Y^n_{s^n})}<\gamma\text{ for all }P_S\in\calP_n(\calS)\right)
\\&\le \bbP\left(\log\frac{(Q_{s^n}W)^n(Y^n_{s^n}|X^n)}{(U_{P_X}Q_{s^n}^nW^n)(Y^n_{s^n})}<\gamma\right)
\\&\le \bbP\left(\log\frac{(Q_{s^n}W)^n(Y^n_{s^n}|X^n)}{(P_XQ_{s^n}W)^n(Y^n_{s^n})}<\gamma+\log|\calP_n(\calX)|\right)\label{eq:terms1c}
\\&\le \rvQ\left(\frac{nI(P_X;Q_{s^n}W)-\gamma-\log|\calP_n(\calX)|}{\sqrt{n\Sigma_n}}\right)\nonumber
\\&\qquad+\frac{K\,T(P_X,P_S,W)}{\Sigma_n^{3/2}\sqrt{n}}\label{eq:terms1d}
\end{align}
where \eqref{eq:terms1c} follows because $U_{P_X}(x^n)\le |\calP_n(\calX)| P_X^n(x^n)$ for all $x^n$, and \eqref{eq:terms1d} follows from Lemma~\ref{lemma:latin_square}, where $\Sigma_n$ satisfies \eqref{eq:Sigma_n_bound}. For the second term in \eqref{eq:dmc_terms}, we have
\begin{align}
&\bbP(Z(X^n,\barX^n,Y^n_{s^n})=0,\,(X^n,Y^n_{s^n})\in\calA)\nonumber
\\&\le \bbP((\barX^n,Y^n_{s^n})\in\calA)\label{eq:terms2a}
\\&=\bbP\left(\log\frac{(P_SW)^n(Y^n_{s^n}|\barX^n)}{(U_{P_X}P_S^nW^n)(Y^n_{s^n})}\ge\gamma\text{ for some }P_S\in\calP_n(\calS)\right)
\\&\le \sum_{P_S\in\calP_n(\calS)} \bbP\left(\frac{(P_SW)^n(Y^n_{s^n}|\barX^n)}{(P_XP_SW)^n(Y^n_{s^n})}\ge\exp\gamma\right)\label{eq:terms2b}
\\&\le \sum_{P_S\in\calP_n(\calS)}  \exp\{-\gamma\}\, \bbE\frac{(P_SW)^n(Y^n_{s^n}|\barX^n)}{(U_{P_X}P_S^nW^n)(Y^n_{s^n})}\label{eq:terms2c}
\\&=|\calP_n(\calS)|  \exp\{-\gamma\}\label{eq:terms2d}
\\&=\frac{1}{M\sqrt{n}}\label{eq:terms2e}
\end{align}
where \eqref{eq:terms2a} follows from the definition of $Z$, \eqref{eq:terms2b} follows from the union bound, \eqref{eq:terms2c} follows from Markov's inequality, and \eqref{eq:terms2d} follows because $\barX^n\sim U_{P_X}$, and so for any $y^n$
\be
\bbE \frac{(P_S W)^n(y^n|\barX^n)}{(U_{P_X}P_S^nW^n)(y^n)}=1,
\ee
and \eqref{eq:terms2e} follows from the definition of $\gamma$ in \eqref{eq:gamma_def}. We may now bound the third term in \eqref{eq:dmc_terms} by writing, for some $s^n$ and $\barx^n$
\begin{align}
&\bbP(Z(X^n,\barx^n,Y^n_{s^n})=0,(X^n,Y^n_{s^n})\in\calA)
\\&\le\bbP(Q_{X^n,\barx^n,s^n,Y^n_{s^n}}\notin\calD_\eta)\label{eq:terms3a}
\\&=  \sum_{\substack{Q_{XX'SY}\in \calP_n(\calX\times\calX\times\calS\times\calY)\setminus\calD_\eta}} \bbP\big( Q_{X^n,\barx^n,s^n,Y^n_{s^n}}=Q_{XX'SY}\big)
\\&\le  \hspace{-.25in}\sum_{\hspace{.2in}Q_{XX'SY}\in \calP_n(\calX\times\calX\times\calS\times\calY)\setminus\calD_\eta}
\hspace{-.65in}\exp\{-nD(Q_{XX'SY}\| P_X\times Q_{X'S}\times W)\}\label{eq:terms3c}
\\&\le (n+1)^{|\calX|^2|\calS|\cdot|\calY|-1} \exp\{-n\eta\}\label{eq:terms3d}
\end{align}
where \eqref{eq:terms3a} holds by the definition of $Z$, \eqref{eq:terms3c} holds by the standard bound on the probability of a type class, and \eqref{eq:terms3d} holds by the polynomial bound on the number of types and the definition of $\calD_\eta$. 

Applying \eqref{eq:terms1d}, \eqref{eq:terms2e}, \eqref{eq:terms3d}, and the definition of $\gamma$ in \eqref{eq:gamma_def} to \eqref{eq:dmc_terms}, we have
\begin{align}
\eps&\le \max_{P_S\in\calP(\calS,\Lambda)}\nonumber
\\&\rvQ\left(\frac{nI(P_X;P_SW)-\log\big[\sqrt{n}\,|\calP_n(\calX)|\cdot|\calP_n(\calS)\,M]}{\sqrt{n\Sigma_n}}\right)\nonumber
\\&+\frac{K\,T(P_X,P_S,W)}{\Sigma_n^{3/2}\sqrt{n}}
+\frac{2\log e}{\sqrt{n}}\nonumber
\\&+2\log(3n|\calS|) (n+1)^{|\calX|^2|\calS|\cdot|\calY|-1} \exp\{-n\eta\}
+\sqrt{\frac{2\ln(3n|\calS|)}{M}}.
\end{align}
Noting that the last two terms are exponentially vanishing (if $M$ is exponentially increasing) and that
\be
\log\big[\sqrt{n}\,|\calP_n(\calX)|\cdot|\calP_n(\calS)=(|\calX|+|\calS|-{\textstyle\frac{3}{2}})\log n+O(1),
\ee
we may rearrange to find
\begin{align}
&\log M\nonumber
\\&\ge \min_{P_S\in\calP(\calS,\Lambda)}\, nI(P_X;P_SW)\nonumber
\\&\quad-\sqrt{n \Sigma_n}\, \rvQ^{-1}\left(\eps-\frac{K\,T(P_X,P_S,W)}{\Sigma_n^{3/2}\sqrt{n}}-\frac{1}{\sqrt{n}}-o(1)\right)\nonumber
\\&\quad-(|\calX|+|\calS|-{\textstyle\frac{3}{2}})\log n-O(1)
\\&\ge \min_{P_S\in\calP(\calS,\Lambda)}\, nI(P_X;P_SW)-\sqrt{nV(P_X,P_S,W)}\, \rvQ^{-1}(\eps)\nonumber
\\&\quad-(|\calX|+|\calS|-{\textstyle\frac{3}{2}})\log n-O(1)\label{eq:final_achievability2}
\\&\ge C-\sqrt{nV_+}\, \rvQ^{-1}(\eps)-(|\calX|+|\calS|-{\textstyle\frac{3}{2}})\log n-O(1)\label{eq:final_achievability3}
\end{align}
where \eqref{eq:final_achievability2} holds by \eqref{eq:Sigma_n_bound} and because moments on $\imath(X;Y)$ may be uniformly bounded for finite $|\calX|,|\calY|$ (cf. \cite[Lemma 46]{Polyanskiy2010}); and where \eqref{eq:final_achievability3} holds by \cite[Lemmas 63 and 64]{Polyanskiy2010}, and because $P_X$ was chosen to be close to $P_X^\star$.
\end{IEEEproof}

\bibliographystyle{IEEETran}
\bibliography{KosutBibs}

\end{document}